\documentclass[prl, twocolumn, superscriptaddress, reprint, aps]{revtex4-1}

\usepackage{float}
\usepackage{empheq}
\usepackage{amsmath}  
\usepackage{amsfonts}
\usepackage{graphicx}   
\usepackage{hyperref}   
\usepackage{cleveref}

\usepackage{bm}
\usepackage{color}
\newcommand{\eq}{\begin{equation}}
\newcommand{\qe}{\end{equation}}
\newcommand{\Fig}[1]{Fig.~\ref{#1}}
\newcommand{\Eq}[1]{Eqn.~\ref{#1}}
\newcommand{\uu}{\mathbf{u}}

\begin{document}

\title{
Conforming nanoparticle sheets to surfaces with Gaussian curvature
}

\author{Noah P. Mitchell}
\email{npmitchell@uchicago.edu}
\thanks{Corresponding author}
\thanks{Equal contribution}
\affiliation{James Franck Institute and Department of Physics, University of Chicago, Chicago, IL, USA}
\author{Remington L. Carey}
\thanks{Equal contribution}
\affiliation{James Franck Institute and Department of Physics, University of Chicago, Chicago, IL, USA}
\affiliation{Cavendish Laboratory, University of Cambridge, Cambridge, UK}
\author{Jelani Hannah}
\affiliation{James Franck Institute and Department of Physics, University of Chicago, Chicago, IL, USA}
\author{Yifan Wang}
\affiliation{James Franck Institute and Department of Physics, University of Chicago, Chicago, IL, USA}
\affiliation{Division of Engineering and Applied Science, California Institute of Technology, Pasadena, CA, USA}
\author{Maria Cortes Ruiz}
\affiliation{James Franck Institute and Department of Physics, University of Chicago, Chicago, IL, USA}
\author{Sean McBride}
\affiliation{James Franck Institute and Department of Physics, University of Chicago, Chicago, IL, USA}
\affiliation{Marshall University, Huntington, WV, USA}
\author{Xiao-Min Lin}
\affiliation{Center for Nanoscale Materials, Argonne National Laboratory, Argonne, Illinois, USA}
\author{Heinrich M. Jaeger}
\email{jaeger@uchicago.edu}
\thanks{Corresponding author}
\affiliation{James Franck Institute and Department of Physics, University of Chicago, Chicago, IL, USA}

\begin{abstract}
Nanoparticle monolayer sheets are ultrathin inorganic-organic hybrid materials that combine highly controllable optical and electrical properties with mechanical flexibility and remarkable strength. 
Like other thin sheets, their low bending rigidity allows them to easily roll into or conform to cylindrical geometries.
Nanoparticle monolayers not only can bend, but also cope with strain through local particle rearrangement and plastic deformation.
This means that, unlike thin sheets such as paper or graphene, nanoparticle sheets can much more easily conform to surfaces with complex topography characterized by non-zero Gaussian curvature, like spherical caps or saddles.
Here, we investigate the limits of nanoparticle monolayers' ability to conform to substrates with Gaussian curvature by stamping nanoparticle sheets onto lattices of larger polystyrene spheres. 
Tuning the local Gaussian curvature by increasing the size of the substrate spheres, we find that the stamped sheet morphology evolves through three characteristic stages: from full substrate coverage, where the sheet extends over the interstices in the lattice, to coverage in the form of caps that conform tightly to the top portion of each sphere and fracture at larger polar angles, to caps that exhibit radial folds. 
Through analysis of the nanoparticle positions, obtained from scanning electron micrographs, we extract the local strain tensor and track the onset of strain-induced dislocations in the particle arrangement. 
By considering the interplay of energies for elastic and plastic deformations and adhesion, we construct arguments that capture the observed changes in sheet morphology as Gaussian curvature is tuned over two orders of magnitude.
\end{abstract}

\maketitle


While any flat thin sheet can easily be rolled into a cylinder, common experience suggests that conforming the same sheet to a sphere is considerably more difficult. 
In order to accommodate the curvature of the sphere, one must fold, cut, or stretch the sheet. 
On surfaces with Gaussian curvature --- that is, curvature in two independent directions, such as on a sphere or saddle --- triangles no longer have interior angles which sum to 180$^{\circ}.$
Conforming a flat sheet tightly to such a surface thus necessarily introduces stresses from stretching or compression. 
If the stresses build up, the material may respond by delaminating or forming cracks, dislocations, or folds~\cite{mitchell_fracture_2017,hure_wrapping_2011,vitelli_crystallography_2006}.  
For applications where initially flat sheets are to conform to arbitrary surface topographies, the ability to cope with Gaussian curvature therefore translates into the ability to bend and deform locally in-plane.

\begin{figure}
	\centering
	\includegraphics[width=\columnwidth]{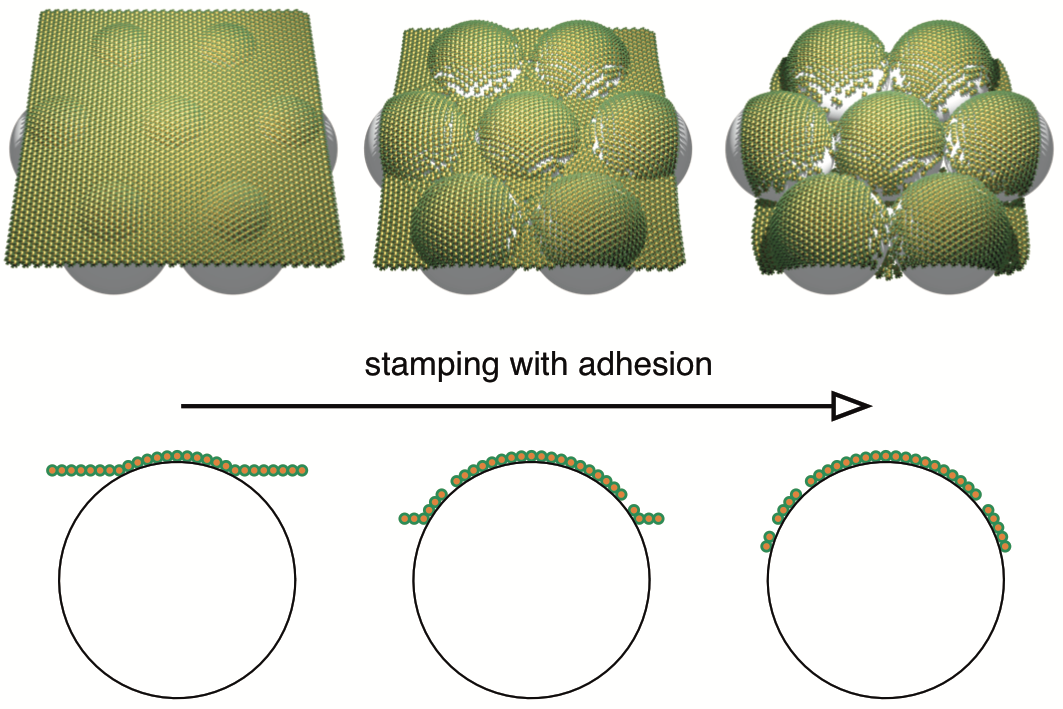}
	\caption{
	\textbf{Nanoparticle sheets conform to highly curved surfaces.} 
	In the situation under study, a preformed nanoparticle monolayer is pressed against a substrate comprised of a lattice of larger spheres. 
	As the sheet is stamped, the nanoparticles become pinned to the substrate spheres.
	The three snapshots (top) are from a simulation of an elastic network. 
 	As the thin sheet conforms to the substrate while experiencing pinning forces, stresses result in broken bonds between nanoparticles.
	}
	\label{fig_setup_sim}
\end{figure}

\begin{figure*}
    \centering
    \includegraphics[width=\textwidth]{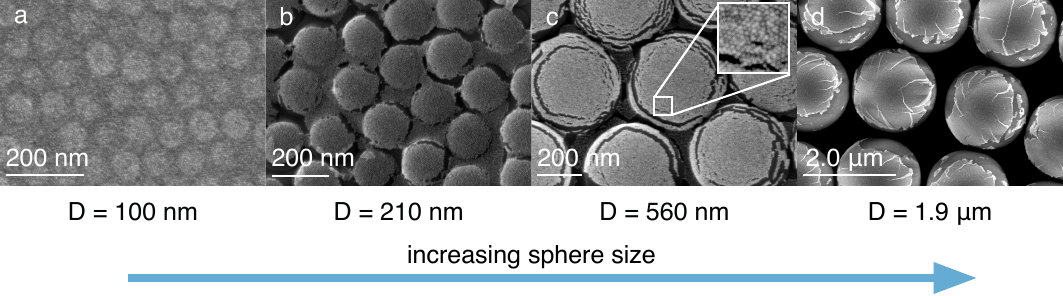}
    \caption{
    \textbf{Sphere size controls the morphology of stamped nanoparticle sheets.} 
    \textit{(a)} At small sphere diameter $D$, the monolayer sheet is able to cover the polystyrene sphere array completely, but does not fully conform to each sphere. 
    \textit{(b-c)} As $D$ increases, the sheets tightly conform to the upper portions of the spheres. However, they no longer bridge the crevices between spheres and instead form azimuthal cracks.
    \textit{(d)} At even larger $D$, sheets buckle out of plane, creating radial folds.}
    \label{fig_morphology}
\end{figure*}

\begin{figure}
	\centering
	\includegraphics[width=\columnwidth]{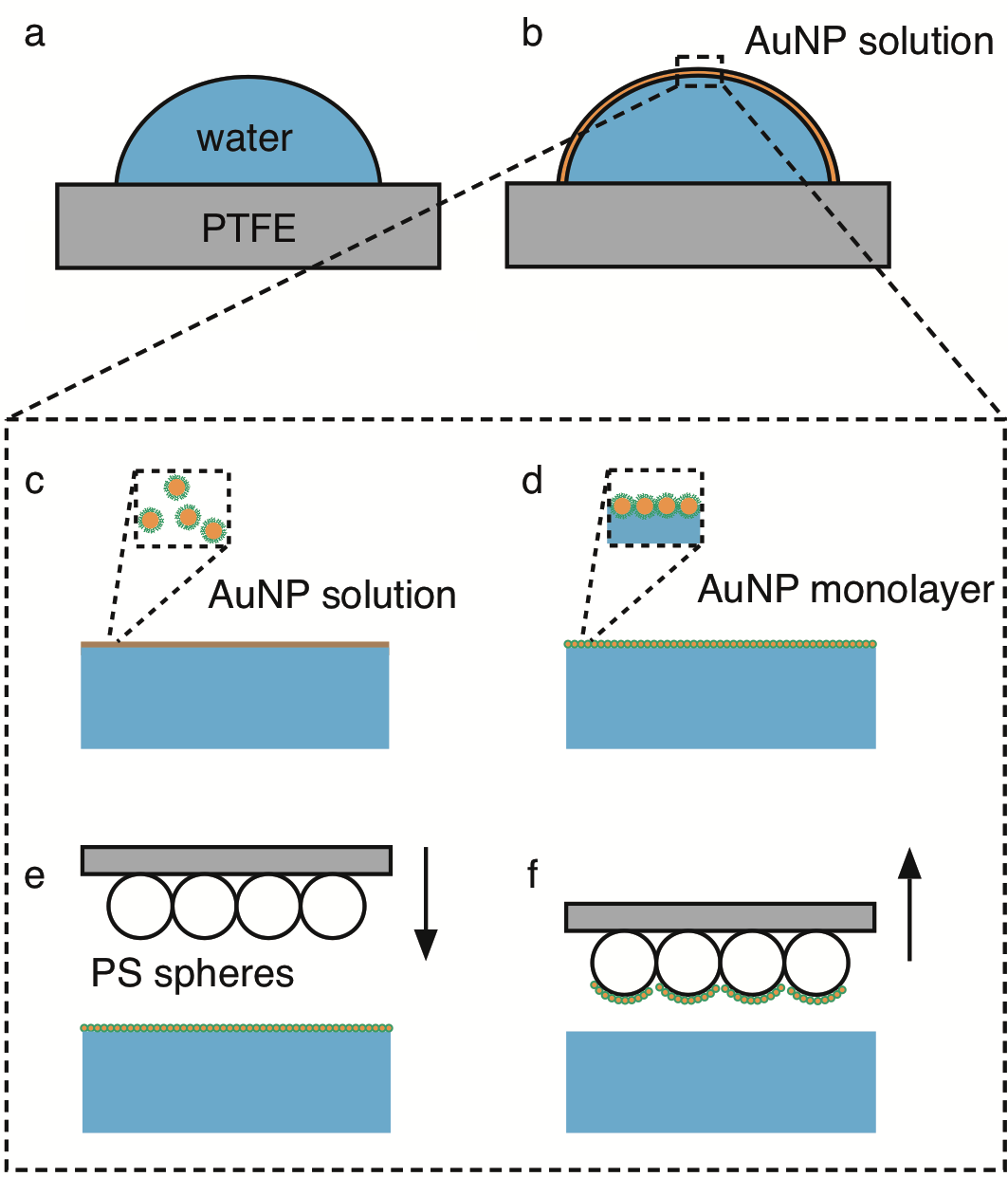}
	\caption{\textbf{Schematic of the experimental procedure for conforming self-assembled gold nanoparticle monolayer sheets to a lattice of polystyrene spheres. }
	\textit{(a-b)} Drying-mediated assembly of a nanoparticle monolayer at the surface of a water droplet.
	\textit{(c-d)} Close-up illustrating the self-assembly of the monolayer at the water-air interface.
	\textit{(e-f)} Stamping a lattice of larger polystyrene (PS) spheres onto the nanoparticle monolayer and peeling it away from the water droplet.}
	\label{fig_procedure}
\end{figure}
Relatively stiff materials such as paper or graphene have difficulty coping with these stresses, and therefore rip or fold instead of conforming to surfaces with Gaussian curvature.
Studies of softer elastic sheets, on the other hand, have led to the understanding of curvature as a tool for patterning defects~\cite{vitelli_crystallography_2006,bausch_grain_2003,irvine_pleats_2010}, 
cracks~\cite{mitchell_fracture_2017}, folds~\cite{paulsen_optimal_2015,yao_planar_2013}, wrinkles~\cite{king_elastic_2012,yao_planar_2013}, blisters~\cite{hure_wrapping_2011}, and even controlling phase transitions to and from the solid state~\cite{meng_elastic_2014,guerra_freezing_2018}.
In this article, we extend these efforts by focusing on a particular material: close-packed nanoparticle monolayers.
These hybrid organic-inorganic materials combine remarkably high Young's modulus (several GPa) with the ability to deform and rearrange locally in a plastic manner.
Furthermore, their versatility has given rise to prospective applications in filters~\cite{he_diffusion_2011}, solar cells~\cite{hau_air-stable_2008}, sensors~\cite{olichwer_cross-linked_2012,he_-situ_2011,saha_gold_2012}, batteries~\cite{tricoli_dispersed_2010}, and beyond due to their optical~\cite{chen_tunable_2008}, electrical~\cite{yang_gold_2016,chen_gold_1998}, and chemical properties~\cite{daniel_gold_2004}.

In nanoparticle monolayers, individual metallic or semiconducting particle cores are embedded in a matrix of interpenetrating ligand molecules that are bound to each core~\cite{nie_properties_2010,dong_binary_2010},
with the organic matrix largely determining the sheet's bulk mechanical properties. 
While these properties have been studied for sheets in planar geometries~\cite{mueggenburg_elastic_2007, gu_tolerance_2017,salerno_high_2014} and for cylindrical, scroll-like structures~\cite{Wang_strong_2015}, the ability of flat sheets to conform to surfaces with Gaussian curvature has received little attention~\cite{rupich_soft_2014}.
Here, we investigate this by stamping monolayers of dodecanethiol-ligated gold nanoparticles onto surfaces formed by lattices of larger polystyrene (PS) spheres.

The situation we address begins with pre-assembled flat sheets that deform as they are stamped against a highly curved surface, as illustrated in~\Fig{fig_setup_sim}.
For nanometer-thin sheets, van der Waals forces generate adhesion that effectively immobilizes the nanoparticles as they come into contact with the substrate.
Furthermore, in contrast to continuum elastic sheets, the discrete nanoparticle lattice allows for the formation and proliferation of defects in addition to straining, folding, and fracturing during the conformation process. 

The effect of strong pinning to the substrate results in strikingly different behavior from that of equilibrium arrangements of interacting Brownian particles on spheres~\cite{irvine_pleats_2010,guerra_freezing_2018}, frustrated equilibrium conformations of macroscopic, continuum elastic sheets~\cite{mitchell_fracture_2017, hure_wrapping_2011}, or non-equilibrium growth of colloidal crystals on spherical interfaces~\cite{meng_elastic_2014}. 
Because the pinned sheet cannot relax to minimize free energy, the effects of geometric frustration build up according to history-dependent, sequential rules.
This sequential adhesion gives rise to qualitatively different stress fields in the sheet and suppresses wrinkling before the appearance of sharp folds.

Depending on the Gaussian curvature, $K$, of the corrugated substrate, which we control by the PS sphere diameter $D$ via $K = 4/D^2$, we find three characteristic stamped-sheet morphologies. 
As seen in~\Fig{fig_morphology}, increasing $D$ leads from sheets that entirely cover the corrugated substrate to sheets that have fractured into caps closely conforming to the top portions of the PS spheres.
Finally, the largest PS spheres yield caps exhibiting radial folds similar to those seen in macroscopic, continuum sheets~\cite{king_elastic_2012}.  
We show that these curvature-dependent morphologies emerge from the interplay between strong pinning to the substrate, elastic energies, and costs for defect formation. This allows us to generate predictions for the conditions required to obtain full coverage and for the limits to which nanoparticle sheets can conform tightly to arbitrarily curved surfaces.

In what follows, we first describe the experiments and resulting sheet morphologies. 
We then provide energy scaling arguments that rationalize the crossovers between stamped sheet morphologies as a function of $D$ or $K$. 
In subsequent sections, we examine each regime in turn and find that detailed measurements corroborate the overall scaling picture.
We directly measure the local strain within the stamped sheets and compare them to simulations of two-dimensional spring networks made to conform to sphere lattices. 
From these measurements and simulations, we determine the onset of finite size effects due to the discrete nature of the nanoparticles.
This analysis provides a correction to the overall scaling picture for small PS sphere sizes and allows us to predict the maximum polar angle up to which the sheet can tightly conform to individual PS spheres without material failure. 

\section{Experimental procedure}

Dodecanethiol-ligated gold nanoparticles were synthesized via a digestive ripening method followed by extensive washing with ethanol and finally dissolution in toluene~\cite{wang_fracture_2014}. 
This process yielded nanoparticles with diameter $5.2\pm 0.3$ nm and ligand lengths $1.7\pm0.3$ nm. 
Nanoparticle monolayers were self-assembled at the surface of a water droplet. 
After depositing a drop ($\sim 150$ $\mu$L) of deionized water onto the hydrophobic surface of a piece of polytetrafluoroethylene (PTFE), 5-7 $\mu$L of the nanoparticle-toluene solution were pipetted around the drop perimeter. 
The solution climbed to the top of the droplet almost immediately, and, as the toluene evaporated, the nanoparticles self-assembled into a close-packed monolayer with a lattice spacing of $7.2 \pm 0.8$ nm (\Fig{fig_procedure}a-d).  
Waiting several hours allowed some of the water to evaporate as well. 
Given the strong pinning of the drop's contact line to the substrate, this evaporation changed the droplet shape from spherical cap to a flattened (not shown in Fig. 3b).

At this stage, a silicon chip coated with a lattice of polystyrene (PS) spheres was gently pressed against the assembled monolayer and peeled away (\Fig{fig_procedure}e,f).
These PS sphere lattices were created by diluting solutions of PS spheres (Bangs Laboratories) by a factor of 100 using deionized water, then depositing  5-7 $\mu$L of the diluted solution onto 25 mm$^2$ silicon chips and allowing them to dry. 
Our experiments used sphere diameters ranging from 100 nm to 1.9 $\mu$m.
Variations in PS sphere sizes increased with their diameters, ranging from a standard deviation of $2\%$ for 100 nm spheres to $12\%$ for 800 nm spheres, while the 1.9 $\mu$m spheres had a standard deviation of $20\%$.
Because the nanoparticle monolayers readily adhere to the PS spheres, the layers delaminate from the water and transfer to the PS spheres, as when inking a stamp. 
These `stamped' monolayers were then imaged using a Carl Zeiss Merlin scanning electron microscope (SEM).

\begin{figure}
\includegraphics[width=230pt]{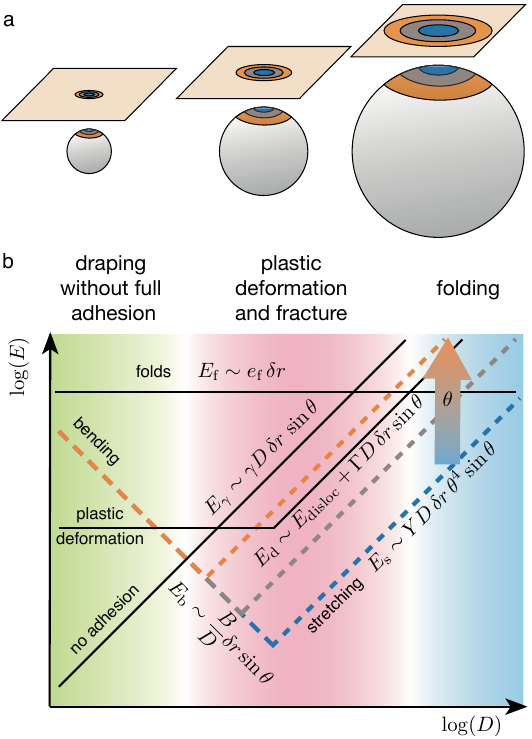}
\caption{
\textbf{Energy scaling captures changes in sheet morphology.}
The interplay of different energy costs provides crossovers from fully covered PS lattices (incomplete adhesion, green region), to plastic deformation (red region), to the formation of localized folds (blue region). 
Each energy is for a nanoparticle annulus of radial width $\delta r$ --- with stretching stiffness $Y$ and bending modulus $B$ --- and a PS sphere of diameter $D$. 
The energy cost of \textit{not} adhering to the PS substrate, $E_\gamma$, grows with the area of the annulus, $\pi D\, \delta r \sin \theta$, and depends on the adhesion energy, $\gamma$.
Similarly, the stretching energy, $E_s$, and the energy of plastically deforming the annulus by dislocation proliferation, $E_\textrm{d}$, likewise grow with the area of the annulus.
The stretching energy also depends strongly on the polar angle, $\theta$, through the strain $\epsilon_{ij}=\epsilon_{ij}(\theta)$ as $E_s \sim D \delta r \theta^4 \sin \theta$, depicted by the offset between colored dashed blue, gray, and orange lines.
The plastic deformation energy, $E_\textrm{d}$, has a minimum set by the energy of unbinding a pair of dislocations, $E_{\textrm{disloc}}$, and the factor $\Gamma$ is a phenomenological constant characterizing the work necessary to plastically deform a unit area of the sheet.
The energy of creating a localized fold, $E_f$, is set by the energy to crease the sheet.
The fold energy per unit length of the fold, $e_f$, depends on the fold angle and microscopic details of the lattice.
}
\label{fig_energy_scaling}
\end{figure}

\section{Monolayer morphology: coverage, cracks, and folds}

SEM imaging revealed that the nanoparticle sheets reproducibly retain their monolayer structure as they are transferred onto the substrate of PS spheres. 
The sheet morphology, however, varies with the size of the PS spheres used. 
For PS diameters $D \approx$ 100 nm, monolayers typically cover the substrate without cracks or folds (\Fig{fig_morphology}a).
For these small $D$, the monolayers do not enter deeply into the crevices between spheres, instead getting pinned at the apex of each PS sphere and bridging the crevices as freestanding membranes.  

Once $D$ becomes larger, the stamped sheets are able to follow the substrate surface topography more closely, creating snugly fitting caps. 
Remarkably, the sheets conform tightly to the PS spheres up to polar angles of 20-30$^\circ$  (measured from the apex of each sphere) without buckling, wrinkling, or creating folds.  
This already indicates behavior quite distinct from that of other thin sheets, such as paper, mylar, polystyrene, or graphene, which invariably generate folds or rip~\cite{hure_wrapping_2011,hure_stamping_2012,paulsen_optimal_2015,Lee_measurement_2008,dudte_programming_2016}.

At larger polar angles, azimuthally oriented cracks appear, which hint at large radial stress as the sheets conform to the PS spheres during the stamping process.
These cracks prevent the sheets from bridging the gap between neighboring spheres (\Fig{fig_morphology}b-c).
For sphere diameters larger than roughly 1 $\mu$m, not only do the sheets tear azimuthally to form caps on each sphere, but also they form localized radial folds to accommodate the mismatch between flat and spherical metrics (\Fig{fig_morphology}d).

The azimuthal cracks in~\Fig{fig_morphology}b-c and the radial folding lines in~\Fig{fig_morphology}d form during the stamping process, in which the monolayers are deformed under vertical pressure to conform against the non-Gaussian topography, as sketched in~\Fig{fig_setup_sim}. 
Once the nanoparticles are in contact with the polystyrene surface, the adhesion immobilizes these local deformations.
For $D$ around 200 nm, portions of the monolayer that did not adhere  to PS spheres tend to tear in the interstices between polystyrene spheres.
For larger $D$, the azimuthal fractures become more pronounced, allowing the interstitial portions of the sheet to recede further down (\Fig{fig_morphology}c).
For the largest sphere sizes ($D \ge 690$ nm), the non-adhering portions may be swept away as the water dewets the chip while it is being pulled off the droplet at the end of the stamping process (\Fig{fig_morphology}d).

\section{Energy scaling}
In this section, we provide a self-consistent rationalization for the observed changes from incomplete adhesion to plastic deformation to folding, using scaling arguments for continuum sheets.
In subsequent sections, we examine each regime in turn and find that detailed measurements corroborate the overall scaling picture presented here, while also providing corrections due to the discrete lattice structure of our sheets.

A simple geometric insight underpins the trend in behavior seen in~\Fig{fig_morphology}. 
On a flat sheet, the circumference of a circle grows in proportion to its radius, $r$. 
On a sphere, however, the circumference of a circle at the same distance $r$ from the sphere's apex grows more slowly due to the Gaussian curvature. 
In other words, when a flat disc of given $r$ is made to conform to the surface of a sphere, it must deform to compensate for the deficit in circumference. 
The sheet must therefore not only bend, but also strain elastically in the form of radial expansion, azimuthal compression, or some combination of the two. 

If the sheets furthermore become pinned to the PS spheres during the stamping process, the nanoparticles attach sequentially one annulus at a time, starting from each sphere's apex (\Fig{fig_energy_scaling}a). 
As successive annuli conform to the substrate, the cost of elastic energy may exceed the energetic costs associated with delaminating, forming defects, ripping apart, or folding. 
To understand the competing energy scales, consider an annulus of nanoparticle sheet with radial width $\delta r$ that has been conformed onto a PS sphere of diameter $D$ to sit at polar angle $\theta$. 
Such an annulus has an area $ \pi D \delta r \sin \theta$ (to zeroth order in strain).   
Conforming this annulus to the sphere requires energies due to bending and stretching, and these conformational energy costs compete with alternative behaviors, such as remaining free-standing instead of conforming, plastically deforming and fracturing, or folding.

\subsection{Energy costs to conform: bending and stretching}
First, conformation of the annulus requires areal bending energy density $\mathcal{E}_\textrm{b} \sim B/D^2$, where $B$ is the sheet's 2D bending modulus. 
The total bending energy in the annulus then becomes $E_\textrm{b} \sim (B/D) \delta r \sin \theta $. 
Here we are neglecting small corrections to this approximation of order $\mathcal{O}(\theta^2)$ (see Supplementary Information). 
Thus, the cost of bending decreases as $D$ grows, as shown by the downward dashed line in the left portion of~\Fig{fig_energy_scaling}b. 

Second, the sheet must also stretch to conform to a sphere.
The total stretching energy, $E_s$, stored in the annulus is proportional to its surface area and the stretching energy density. 
This stretching energy density, $\mathcal{E}_s$, is a quartic function of polar angle on the sphere, $\mathcal{E}_s \sim Y \theta^4$, as shown in the Supplementary Information. 
Therefore, the cost of stretching increases linearly with $D$, but the magnitude depends sensitively on the polar angle: $E_s \sim Y D  \delta r \, \theta^4 \sin \theta$.
While~\Fig{fig_energy_scaling}b omits linear and sublinear dependence on $\theta$ for clarity, this strong dependence of $E_s$ on polar angle is shown by the rising dashed lines. 
The changing colors (blue, gray, orange) denotes that, for a given sphere size $D$, the stretching energy in an annulus grows rapidly with polar angle.

We emphasize that the stretching energy scaling in our sheets strongly contrasts from the well-studied case of equilibrated sheets conformed to a sphere, in which the energy density \textit{decreases} quadratically with polar angle, $\theta$ for small $\theta$ (see Supplementary Information).
This difference highlights the distinct character of sequential adhesion to a substrate seen in our system.

\subsection{Alternatives to elastic conformation: avoiding adhesion, plastic deformation, and folding}

These elastic energies compete with the possibility of adopting alternative behaviors. 
Instead of elastically bending and stretching to conform, the sheet may only partially conform to the sphere, or it may plastically deform, rip apart, or form folds.

While stretching and bending cost energy, the adhesion process can \textit{relieve} energy as well, since it replaces two interfaces (nanoparticle-air and air-PS) with a single one (nanoparticle-PS).
This replacement relieves energy in proportion to the area of adhered material, so there is a fixed areal energy density $\mathcal{E}_\gamma$ relieved by adhering to the PS sphere.
For the annulus, this translates into a total cost of \textit{not} adhering to the substrate, $E_\gamma \sim \gamma D  \delta r \sin \theta$, that increases linearly with $D$.

While the stretching energy scales as $E_s \sim Y D  \delta r \, \theta^4 \sin \theta$, the energy cost $E_\textrm{d}$ of relieving stress through plastic deformation of the annulus scales similarly with sphere diameter, but has a far weaker scaling in $\theta$: $E_\textrm{d} \sim \textrm{max}(E_{\textrm{disloc}}, \, \Gamma D \delta r \sin \theta )$, where $E_\textrm{disloc}$ is the energy of unbinding a single pair of  dislocations and $\Gamma$ is a phenomenological factor capturing the work required to damage a unit area of the material. 
The minimum possible energy to create the first defect pair, $E_{\textrm{disloc}}$, sets the lower cutoff that freezes out defect proliferation at small $D$. 
$E_{\textrm{disloc}}$ is determined by the core energy of a dislocation and the elastic cost of deforming the portion of sheet surrounding the dislocations, which depends on microscopic features of the lattice. 
Finally, the energy cost for creating a fold in the sheet, $E_f$, increases only with the fold length ($E_f \sim e_f \delta r$, where $e_f$ is the fold energy per unit length) and thus is independent of $D$.

\subsection{Three regimes arise from energy scaling}
\Fig{fig_energy_scaling}b represents these energy scaling relations schematically. 
Throughout this figure, linear and sub-linear dependences on the polar angle $\theta$ are suppressed for clarity. 
In particular, the adhesion and bending energies grow as $\sin\theta$, and we omit this dependence. 
Conversely, we do include the strong $\theta$ dependence of the stretching energy, and illustrate this strong dependence by the colored dashed lines.

From this scaling we infer that for sufficiently small sphere sizes (or, equivalently, large Gaussian curvature), the lowest cost will be incurred by incomplete adhesion, as this causes the least distortion in the flat sheet. 
The green region in~\Fig{fig_energy_scaling}b represents this regime, which corresponds to the experimental results in~\Fig{fig_morphology}a.

For larger sphere sizes, bending becomes energetically cheaper than not adhering. 
However, in order to conform tightly to the sphere, the monolayer needs to not only bend, but also stretch or compress. 
For annuli at small polar angles $\theta$, this elastic energy cost can be negligible, but as $\theta$ grows for a given $D$, the cost will eventually exceed the penalty for creating defects. 
As a result, beyond some critical polar angle $\theta_c$, plastic deformation in the sheet will cause a proliferation of dislocations. We expect that the formation of cracks follows as a result of this defect formation, along with the tension that remains while defects are formed. 
Since the in-plane stretching is tensile along the radial direction, as we will see, cracks open up along the azimuth, perpendicular to the radial tension. 
This regime is represented by the red region in~\Fig{fig_energy_scaling}b and corresponds to the experimental results in~\Fig{fig_morphology}b and c.

For the largest PS sphere sizes, yet another crossover occurs due to the difference in scaling between the costs for either elastic stretching or plastic deformation, which increase linearly in $D$, and the costs of forming localized folds, which is independent of $D$. 
This is the regime shown in blue in~\Fig{fig_energy_scaling}b, corresponding to~\Fig{fig_morphology}d. 
Because the energy cost for fold formation lies below that of plastic deformation in the blue regime, the first response as strains build up will be to form folds rather than the proliferation of dislocations.

This energy scaling captures all three regimes of stamped nanoparticle sheet morphology seen in~\Fig{fig_morphology}.  
We note that this framework operates in the continuum limit.
Additionally, our picture assumes that chemical properties of the polystyrene do not vary with PS sphere size, an effect that could alter the adhesion energy in~\Fig{fig_energy_scaling}b.
Nevertheless, the essential features are supported by quantitative comparisons with experiments and simulations given in the following sections. 

In the remaining sections, we discuss in more detail each of the mechanical responses of the flat sheets to the enforced geometric mismatch: bending, stretching, dislocation proliferation, crack formation, and folding.

\section{Bending and adhesion}
The crossover from incomplete adhesion to full adhesion with plastic deformation occurs in our experiments for PS spheres with diameters $D\approx 200$ nm.
This crossover enables an estimate of the bending rigidity in nanoparticle membranes.

Near the apex of the sphere, the two-dimensional bending energy density of a thin plate in plane stress is~\cite{landau_chapter_1986}
\begin{equation}
\mathcal{E}_{\textrm{b}} \approx \frac{4 B (\nu +1)}{D^2}.
\end{equation}
We take the Poisson ratio to be $\nu=1/3$, the value for a triangular lattice of spring-coupled nodes, in accordance with the measured value for nanoparticle sheets~\cite{Kanjanaboos_strain_2011}.
We take an average radius of curvature of $D/2 \approx 100$ nm for the crossover. 



At the small-sphere crossover between incomplete adhesion and plastic behavior, we should expect the bending energy to match the adhesion of the nanoparticle sheet with polystyrene. 
Using the result of Ref.~\cite{girifalco_theory_1957}, we estimate the adhesion energy from the surface tensions of dodecane (21 mN/m) and water (72 mN/m), the surface energy of solid polystyrene ($\sim$ 42 mN/m)~\cite{Wu_calculation_2007}, and the molar volumes of each.
The result is an adhesion energy of $\gamma_{\textrm{PS}} + \gamma_{\textrm{dodecane}} - \gamma_{\textrm{PS},\textrm{dodecane}}  \approx$ 60 mN/m.
We expect that the bending energy, $\mathcal{E}_{\textrm{b}}$, matches this value at the crossover. 
This gives a bending modulus for the nanoparticle sheets of $B \approx 4.5 \times 10^{-16}$ Nm.

From this we may deduce a lower bound on the effective thickness $t_{\textrm{eff}}$ of the sheet, which can deviate from the physical thickness due to the non-continuum nature of the material~\cite{Wang_strong_2015}.
The bending modulus is related to $t_{\textrm{eff}}$ via $B = Y t_{\textrm{eff}}^2 /12(1-\nu^2)$. 
Here the 2D stiffness $Y=E t$ is the product of Young's modulus $E$ and  physical thickness $t$.
If we assume $E \sim 3 $ GPa, as is appropriate for fully dried monolayers~\cite{he_fabrication_2010,mueggenburg_elastic_2007}, we obtain $t_{\textrm{eff}} \approx 14$ nm, about 60\% larger than the physical thickness of $t \approx (d_{\textrm{Au NP}} + 2 \times \ell_{\textrm{ligand}})$ nm $=8.2$ nm. 
However, we expect that  during the stamping process there is residual water embedded in the ligand matrix.
The presence of water molecules in the matrix has been shown to drastically affect the elastic properties, reducing elastic moduli by potentially several orders of magnitude~\cite{wang_thermomechanical_2017,griesemer_role_2017}.
 Such decrease in $E$ then implies an increase in $t_{\textrm{eff}}$, possibly up to around 10$t$ as observed for dried monolayers~\cite{Wang_strong_2015}.


The crossover from incomplete adhesion on small spheres to tightly conforming to larger spheres is reminiscent of the crossover in a thin sheet's `bendability', which is the ratio of tensile to bending forces, $T W^2/B$, where $T$ is the tension at the edge of a sheet of width $W$ due to in-plane stretching or interfacial forces~\cite{king_elastic_2012}.
If we consider the case where $W\sim D$, so that the sheet covers the same proportion of the sphere for different sphere sizes, then as the PS sphere size increases, so too does the bendability of the sheet.
Our system differs from these recent studies of comparably stiff sheets, however, because of the strong pinning of the nanoparticle sheet to the substrate. 
The apparent force imbalance in the stretching of the sheet measured in simulations shows that adhesion enables a disproportionate increase in radial tension, at a rate faster than long-range elasticity would allow (Supplementary Information \Fig{fig_imbalance}). 
Specifically, adhesion supplies a tension which offsets the imbalance of in-plane stresses, $\partial_r (r\sigma_{rr}) - \sigma_{\phi\phi}$. 
While this quantity would vanish without pinning, here the stress imbalance grows as $\theta^2$ for small to moderate polar angles (see Supplementary Information).

\begin{figure}
\includegraphics[width=\columnwidth]{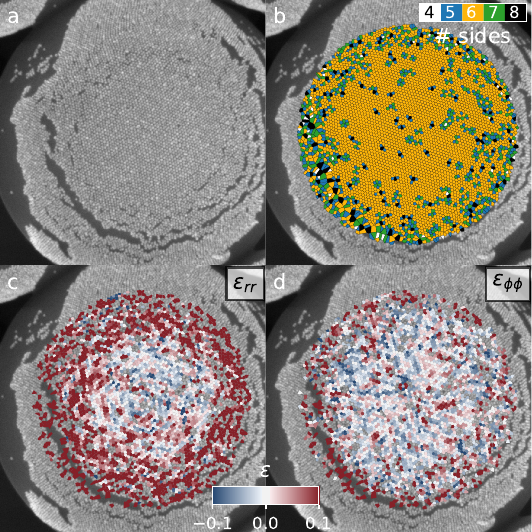}
\caption{
\textbf{Identification of defects and extraction of the local strain tensor.}
\textit{(a)} Nanoparticles are identified in the original SEM image. 
\textit{(b)} Using a Voronoi tessellation, we enumerate the neighbors of each nanoparticle. 
For each nanoparticle with six neighbors, comparing the Voronoi cell to a regular hexagon lying on the tangent plane of the sphere yields the strain tensor. 
To restrict the analysis to elastic deformations, we omit particles whose Voronoi cell is deformed well beyond the elastic limit of the material, keeping only hexagons whose perimeter to surface area ratio, $s \equiv P / \sqrt{A}$, satisfies $s < s_{\textrm{cutoff}} = 3.8$. 
\textit{(c-d)} 
The radial strain in the sheet, $\epsilon_{rr}$, increases with distance from the apex, while azimuthal strain, $\epsilon_{\phi\phi}$, does not.
}
\label{fig_strain_example}
\end{figure}

\begin{figure}
\includegraphics[width=\columnwidth]{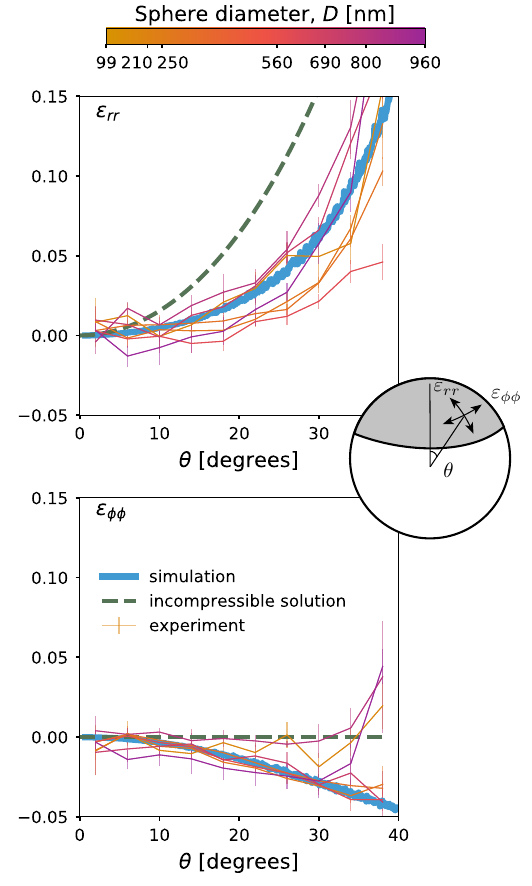}
\caption{
\textbf{Strain analysis shows qualitative agreement between experiments and simulations.} 
Data from nanoparticle sheets on 62 imaged PS spheres of different diameters reveals that the radial strain, $\epsilon_{rr}$ increases with polar angle, while the azimuthal strain, $\epsilon_{\phi\phi}$, is compressive and comparatively small.
The incompressible solution does not fit as well to the data, showing that nanoparticle sheets behave elastically.
}
\label{fig_strain_plots}
\end{figure}


\section{Strain analysis}
During the stamping process, the first contact between the nanoparticle sheet and a PS sphere occurs at the sphere's apex, $\theta=0$, where the sheet will be pinned. 
Subsequent annuli of the sheet will need to strain or undergo plastic deformation in order to conform tightly to the surface of the PS sphere, but once this has occurred, these annuli also will become pinned to the polystyrene.  
This means that we can obtain information about the local strain by using the individual nanoparticles as markers and extracting differences in their average spacing along a sphere's surface.  
Given the random disorder inherent already in the flat sheets, this procedure requires ensemble averages over several different imaged PS spheres for statistically relevant results.


\subsection{Image analysis}
To study the strains and defect densities of nanoparticle sheets, we use a custom image analysis routine on each SEM image to identify the nanoparticle locations and to identify the nearest-neighbor connectivity of the nanoparticle lattice~\cite{crocker_methods_1996}. 
We bandpass each image in two steps: first convolving it with a Gaussian (whose parameters include nanoparticle characteristics such as lattice spacing) and then convolving the result with a boxcar function. 
Subtracting the two gives a high-pass-filtered image from which we extract particle positions.

A Delaunay triangulation provides the lattice topology and the nearest neighbors for each particle.
Defects in the lattice are particles with fewer than six or greater than six neighbors (disclinations), and pairs of oppositely signed disclinations form dislocations (for example, a 5-7 disclination pair).
\Fig{fig_strain_example}b shows an example Voronoi tesselation of a triangulated nanoparticle sheet draped on a 690 nm diameter PS sphere. 
The Delaunay triangulation also enables a direct measurement of the local strain tensor, $\epsilon_{ij}$. 
For particles with exactly six neighbors, we measure the displacements of its neighbors from a regular hexagon with bonds of unit length. 
In this step, we account for the non-planar geometry of the substrate by computing displacements only in the tangent plane to the underlying PS sphere.
By comparing each triad of the central particle and two adjacent neighbors to an undeformed reference triangle, we obtain a strain tensor for that triad of nanoparticles. 
For each particle that is not a defect, the average strain field of its six shared triangles represents a measure of local strain.  
This strain measurement is well-defined only for particles that have six nearest neighbors --- that is, those particles which do not form topological defects in the lattice.

Identifying the center of each substrate sphere by fitting their profile to a circle, we rotate the strain field $\epsilon_{ij}$ into polar coordinates $(\epsilon_{rr}, \epsilon_{r\phi}, \epsilon_{\phi\phi})$ and average annular bins (i.e., bins of $\phi_{i} <\phi <\phi_{i+1}$) to obtain curves for $\epsilon_{rr}(\theta)$ and $\epsilon_{\phi\phi}(\theta)$ as a function of polar angle on a sphere.
Typical results are shown in~\Fig{fig_strain_example}c-d.
\Fig{fig_strain_plots} shows strain curves averaged over several spheres and images for each sphere size. 
To further reduce noise from voids and defects, we also omit particles whose Voronoi cells are deformed well beyond the elastic limit of the material. 
Specifically, we enforce a cutoff in the shape parameter $s$, defined as the ratio of the perimeter of the hexagon to the square root of its surface area, $s \equiv P / \sqrt{A}$. 
Here, we use the cutoff $s < s_{\textrm{cutoff}} = 3.8$, which removes outliers subject to more than 17\% pure shear.



\Fig{fig_strain_plots} shows the average strain tensor components as a function of polar angle for different sphere sizes. 
The analysis indicates that the sheet's radial tension grows substantially, while the strain along the azimuth of the PS sphere is weakly compressive. 
The shear strain averages to zero, as predicted by the symmetry of the spherical geometry, with variations in the measured mean shear of $< 1 \%$.
As mentioned above, the nanoparticle sheets' inherent disorder creates a distribution of strain component values for each binned annulus. 
These distributions have a standard deviation of $\sim 10\%$ strain --- significantly larger than the strains themselves for all but the largest values of $\theta$ considered. 
By averaging the strains in annular bins on each PS sphere and by performing ensemble averages over different spheres, the disorder on the scale of individual nanoparticles is largely averaged out.
As~\Fig{fig_strain_plots} shows, these ensemble-averaged data can show  quantitative differences as the PS sphere diameter $D$ is varied. This likely is due to slight, unavoidable variations in the sample preparation conditions. 
However, within this variability we find no clearly discernible trends as a function of $D$.
Considered in aggregate, these data can therefore be used for qualitative comparison with models, as we discuss next.   

\begin{figure}
\includegraphics[width=\columnwidth]{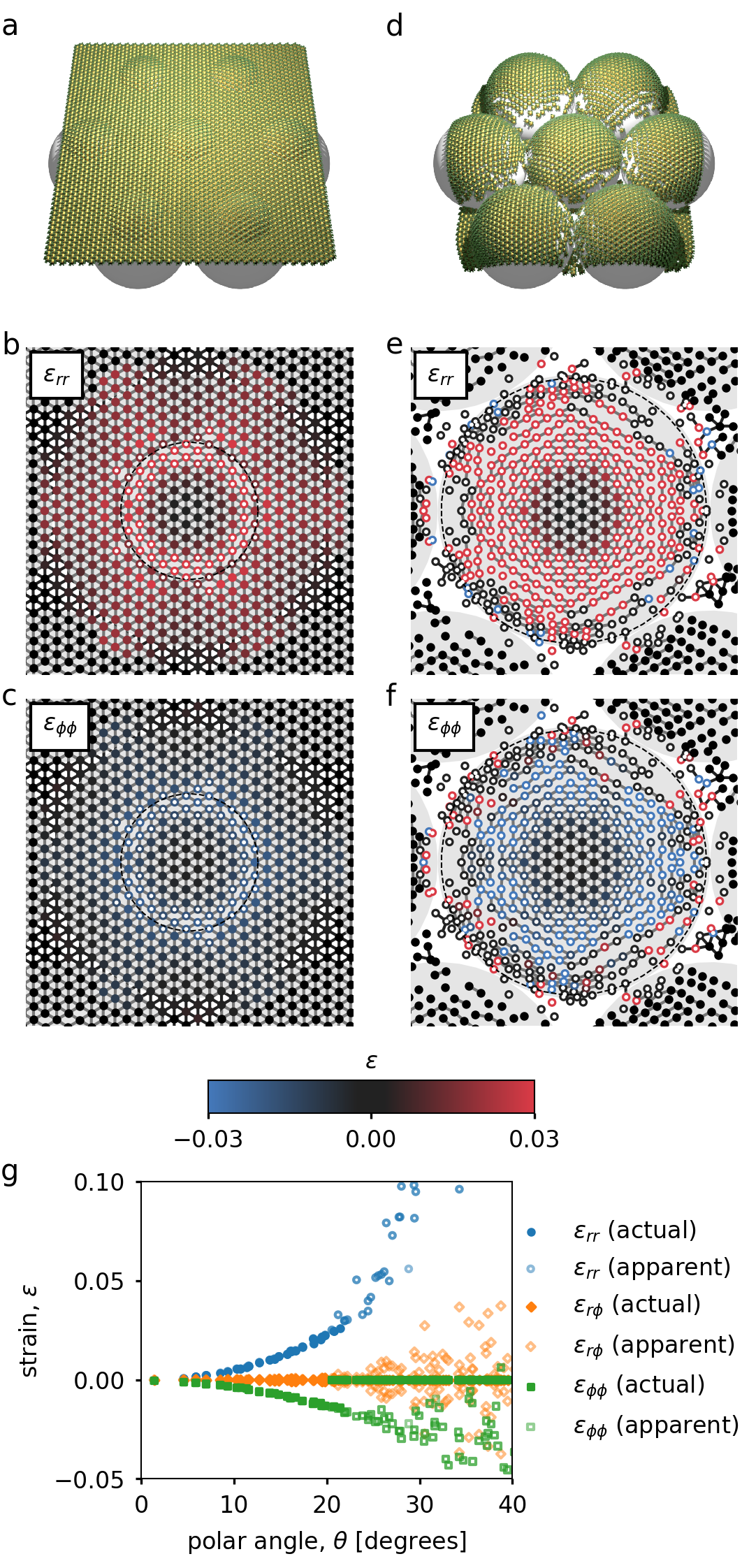}
\caption{
\textbf{Simulations of spring networks with bond breaking reproduce behavior seen in experiment.} Spring networks were made to conform to a lattice of spheres, as in~\Fig{fig_setup_sim}. 
Bonds with $\ge 3\%$ strain are removed at each time step, mimicking bond breakage. 
\textit{(a-c)} As a flat, triangular spring network is pressed against an array of spheres, each node is immobilized upon contact with a substrate sphere. 
As the network conforms, strains build up, leading to bond breaking for polar angles larger than $\theta ~\sim 23^{\circ}$. 
Particles with severed bonds are colored white at their centers in the strain images.
\textit{(d-f)} Layers of bonds continue to adhere to the substrate with many radial bonds broken.
\textit{(g)} Though the actual strains in the network's springs do not exceed $3\%$, the apparent strain inferred from the placement of nanoparticles continues to increase in the damaged annuli.
}
\label{fig_simulation}
\end{figure}

\subsection{Spring network simulations}

To gain insight into the elastic behavior during the stamping process, we model the nanoparticle sheet as a flat, triangular spring network. 
Simulations of such networks pinned to a lattice of spheres reproduce the trends in strain observed in the experiments (\Fig{fig_strain_plots} and Supplementary Videos 1-3).

The simulations proceed by minimizing the free energy of a triangular spring network at each time step using a conjugate gradient method as we deposit the network onto a lattice of spheres. 
Whenever a node of the spring network makes contact with a substrate sphere, we irreversibly pin that node to the point of contact for the remainder of the simulation.
Increasing the radii of the substrate spheres with respect to the bond length by a factor of two (and, proportionately, scaling the number of nanoparticles by a factor of four) gave virtually identical results for the strain plots given in~\Fig{fig_strain_plots} and~\Fig{fig_simulation}, indicating that the simulations are representative of the continuum limit. 
Study of the finite size scaling shows that the strain curves deviate significantly from the continuum limit only for substrate sphere sizes below $D\lesssim 10 a$, where $a$ is the lattice spacing (see Supplementary Information \Fig{fig_finsize}).

In the simulations, a sheet began at a distance $R=D/2$ above the plane containing the centers of the substrate spheres, each of diameter $D$. 
The network was then lowered in small increments $(0.001 D)$ and the free energy was minimized for that configuration, subject to the constraint that all particles (nodes of the spring network) must lie in the common membrane plane or on a sphere, whichever is higher in the $z$ dimension. 
For each step, a sequence of random kicks were applied to each node to escape local minima in the energy landscape. 
At the end of the relaxation process, nodes in contact with a substrate sphere --- that is, within a small threshold of $10^{-5}a$, where $a$ is the rest bond length (lattice spacing) --- are marked as immobilized for the remainder of the simulation.

As shown by the blue curves in~\Fig{fig_strain_plots}, as well as in Supplementary Video 3, these simulations of perfectly elastic triangular networks show similar behavior in both $\epsilon_{rr}$ and $\epsilon_{\phi\phi}$ as a function of polar angle on the underlying sphere. 
As the membrane begins to conform to the sphere lattice, pinning ensures that the apex of the sphere experiences negligible strain, as expected.
The radial stress increases quadratically, while a compressive azimuthal stress builds up more slowly.
The deviation of $\epsilon_{\phi\phi}$ between experiment and simulation at large $\theta$ is due in part to the material failure and plastic deformation of the actual sheets, which is suppressed in the simulations we show in~\Fig{fig_strain_plots} (see also Supplementary Video 3).

We note that in experiment, the nanoparticle membrane may not be perfectly flat in the interstices of the PS spheres, as the pressure of the water during stamping may push the sheet into the interstices. Modifying the simulation geometry to enforce an indentation of the sheet into the interstices of the PS lattice has only a weak effect leading to somewhat elevated strains in the final, pinned state without changing the qualitative strain behavior (Supplementary Information~\Fig{fig:interstices}).


\subsection{Comparison with incompressible solution}
Considering the limit in which the nanoparticle sheet is incompressible allows for a useful point of reference against which we can compare the iterative adhesion of nanoparticle annuli.
The strains required to conform to the substrate in this limit are indicated by the green dashed line in~\Fig{fig_strain_plots}.
Namely,
\begin{equation}
\epsilon_{rr} = \sqrt{\frac{R^2}{(R^2 - r^2)}} - 1 ,
\end{equation}
where $R=D/2$ is the radius of the PS sphere, while $\epsilon_{\phi\phi} = 0$ due to incompressibility.
All data, whether experimental or simulation-based, lie below this solution for $\epsilon_{rr}$. 
This clearly indicates compressible behavior of our nanoparticle sheets.

\subsection{Azimuthal cracks in simulations}
The material cannot stretch elastically without bound: sufficiently large strains will plastically deform the sheet, severing bonds between nanoparticles to form cracks or dislocations.
Indeed, the radial strains seen in~\Fig{fig_strain_plots} greatly exceed the critical strain for failure in flat nanoparticle membranes~\cite{wang_fracture_2014}.
While we will consider plastic deformation in the next section, we note that introducing failure into the spring network simulations generates qualitatively similar morphologies to those seen in experiment.
\Fig{fig_simulation} and Supplementary Videos 1 and 2 demonstrate that introducing a nominal breaking strain of $3\%$ leads to the formation of partially intact annuli separated by azimuthal cracks. 
In~\Fig{fig_simulation}g, we show both the strains of particles with all original bonds intact (closed markers) as well as the `apparent' strain (open markers) resulting from triangulating the point pattern and including all particles with six nearest neighbors, regardless of whether the bonds connecting them have severed.   
This gives strains that remain qualitatively similar to those seen in experiment, with increased scatter in the apparent strains frozen into the broken regions pinned to the substrate.


\section{Plastic deformation}

Given that a flat nanoparticle lattice forms a close-packed array of hexagons, any particles that do not have six nearest neighbors are defects. We record the location of each defective particle and its number of nearest neighbors.
\Fig{fig_strain_example}b shows the Voronoi tessellation of one representative lattice overlaying the original SEM image. 
Each yellow site corresponds to a nanoparticle having six nearest neighbors (i.e., a hexagon), while defects are colored white, blue, green, and black for coordination numbers of $z=4,5,7,$ and $8$, respectively.

As the sheet begins to respond with plastic deformation, dislocations proliferate in the material. 
The density of dislocations correspondingly increases with polar angle on a sphere, as can be seen in~\Fig{fig_strain_example}b.
We observe that azimuthal cracks form only beyond the point of dislocation proliferation, which suggests that the material yields plastically before cracks coalesce.

\subsection{Formation of dislocations}

\begin{figure}
\includegraphics[width=\columnwidth]{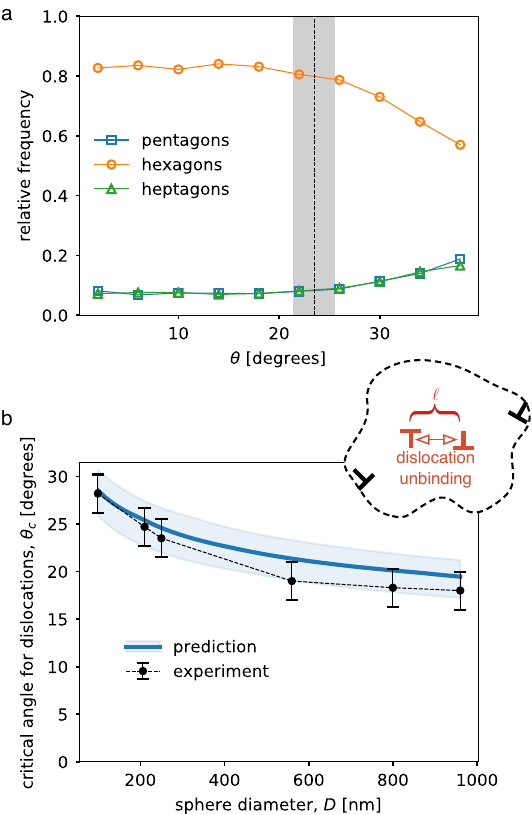}
\caption{
\textbf{Strain-induced defects in the nanoparticle sheets reveal non-continuum behavior.}
\textit{(a)}
The proliferation of defects results in increasing dislocation frequency (and correspondingly, to a decreasing frequency of hexagons) as a function of polar angle, $\theta$. 
An example of the angle-dependence of defect densities is shown for nanoparticle sheets conformed to 250 nm PS spheres. 
Here, a crossover appears near $\theta_c \sim 24^{\circ}$.
\textit{(b)} For small sphere diameters, the characteristic angle for defect proliferation deviates from its continuum value, with smaller PS spheres triggering the formation of defects at larger polar angles. 
An idealized prediction for the energy of a single defect provides a rough estimate for the critical angle (blue curve with blue band denoting the uncertainty from the spread in measurements of the defect density). 
Data for the smallest sphere diameters included only sheets stamped on isolated spheres, not sheets which cover close-packed PS lattices.
 }
\label{fig_dislocations}
\end{figure}

The scaling arguments presented in~\Fig{fig_energy_scaling}, which operate in the continuum limit, predict that plastic deformation should be favorable at a critical angle independent of sphere diameter $D$. 
In our experiments, however, we observe an increase in the polar angle at which dislocations appear for the smallest PS sphere sizes, shown in~\Fig{fig_dislocations}.
This observation implies that the discrete structure of the nanoparticle monolayers can be important in determining the details of their mechanical behavior.
The continuum limit description of~\Fig{fig_energy_scaling} does not include microscopic details, and therefore predicts a size-independent critical angle for the onset of plasticity. 
If the discrete structure of the sheet comes into play, we expect a correction to this picture to appear at small sphere sizes, where the lattice spacing is a non-negligible fraction of the system size.

As expected, the most prominent types of strain-induced defects in the nanoparticle arrangement are dislocations --- i.e., pairs of Voronoi cells with 5 and 7 sides.
\Fig{fig_dislocations}a shows a representative measurement of the crossover from low to high defect density as a function of polar angle, $\theta$. 
These data were obtained from ensemble averages over Voronoi tessellations such as that shown in~\Fig{fig_strain_example}b.
For each PS sphere diameter $D$, we identify a characteristic angle at which the number of defects begins to grow significantly (black dashed line in~\Fig{fig_dislocations}a).
This analysis leads to the black data in~\Fig{fig_dislocations}b, which shows the characteristic angle as a function of $D$. 
This angle approaches a constant value consistent with scale-invariance in the continuum limit of large PS sphere sizes, where the nanoparticle lattice spacing becomes irrelevant. 
However, we observe an increase in the angle for the smallest PS sphere sizes. 
This observed variation in the onset of dislocation proliferation suggests that the discrete nature of the lattice becomes important for small $D$.
If we approximate our sheet as a locally flat, two-dimensional lattice, each dislocation pair costs an elastic energy~\cite{Weertman_elementary_1992}
\begin{equation}\label{eq_dislocation_unbinding}
E_{\textrm{disloc}} \approx \frac{\mu a^2}{2 \pi  ( 1 - \nu)} \ln \left( \frac{\ell}{a} \right),
\end{equation}
where $Y$ is the sheet stiffness, $\nu$ is the Poisson ratio, $\ell$ is the final distance between the unbound dislocations, and \textit{a} is the lattice spacing. 
We assume the elastic core energy to be small compared to the elastic energy in the deformed sheet, with the understanding that Equation~\ref{eq_dislocation_unbinding} represents a lower bound.
Below, we consider $\ell \approx 1/3\sqrt{\rho}$, as illustrated in the inset of~\Fig{fig_dislocations}. 
Here, $\rho$ is the density of dislocations (so that $\rho^{-1}$ approximates the area of a patch whose elastic deformation is dominated by the dislocation's presence).
Note that we expect this elastic energy to be felt predominantly in regions of the material which are not already pinned to the underlying substrate.

In order to find a lower bound for the critical angle at which defects may appear, we compare the dislocation unbinding energy (Equation~\ref{eq_dislocation_unbinding}) with the stretching energy for the sheet to conform to a sphere.
Using the results from spring network simulations, we equate the stretching energy available in an annulus of width chosen to be $\delta r = a$ with the unbinding energy of Equation~\ref{eq_dislocation_unbinding}.
This gives the blue solid line in~\Fig{fig_dislocations} for $\ell = (3\sqrt{\rho})^{-1}$, with the blue band denoting the range of results given the standard deviation of measurements for $\rho$ across sheets on all PS spheres included in the analysis.
As seen by the width of the blue band, the prediction is moderately sensitive to the assumed distance that the unbound dislocation travel apart in their creation.
We measured the dislocation density, $\rho$, from the relative frequency of dislocations at $\theta=0$ in experiments.
Despite the approximate nature of the derivation, the prediction lies within our experimental uncertainty for changes in the choice of $\delta r$ by up to a factor of three, and the agreement in the shape of $\theta_c(D)$ is notable.

\subsection{Formation of azimuthal cracks}
Another response to the buildup of strain is to form cracks in a material. 
This irreversible deformation relieves elastic energy by severing bonds between nanoparticles. 
We find that, for PS sphere sizes above 210 nm, nanoparticle sheets generally form azimuthal cracks such as those seen in~\Fig{fig_morphology}c and~\Fig{fig_strain_example}. 

From a geometric standpoint, projecting an annular strip of inner diameter $D\theta_0$ from a flat disk onto a sphere of diameter $D$ involves less azimuthal compression if the annulus is placed at a polar angle $\theta_1 > \theta_0$.
This fact is reflected in our experiments and simulations, with radial strain building up with increasing polar angle.
Once the radial strains are sufficient to rip apart bonds to form azimuthal cracks, we expect that as the next portion of the membrane drapes onto the sphere, it is energetically favorable to adhere to a location further down, where $\theta_{\textrm{adhere}} > \theta_{\textrm{rip}}$. 
The result is a portion of uncovered PS sphere between $\theta_{\textrm{rip}}$ and $\theta_{\textrm{adhere}}$, i.e., an azimuthal crack imprinted on the spherical substrate.

\section{Formation of folds at large sphere sizes}

For the largest PS sphere sizes, the caps formed by the adhering nanoparticle sheets are large enough that radially oriented folds can be observed (\Fig{fig_morphology}d). 
Such folds provide an alternate mechanism to map circles in the plane to circles on a sphere while minimizing radial tension and azimuthal compression. 
Localizing elastic energy into folds relieves the stretching in intervening patches. 
At the same time, because of the very high curvature in one dimension at the fold (which we expect to be comparable to the inverse lattice constant, $a^{-1}$), the energetic barrier to fold formation is larger than the bending energy by a factor $\sim D^2 / a^2$, implying that the cost of having a fold in an annulus of fixed width, $\delta r$, does not vary with sphere diameter $D$.
This means that, for sufficiently large $D$, where the elastic cost of stretching grows higher and higher, fold formation is no longer frozen out (\Fig{fig_energy_scaling}).

In previous studies of folding that subjected thin sheets to uniaxial compression or out-of-plane deformation, folds often span the whole system~\cite{pocivavsek_stress_2008, holmes_draping_2010,kim_hierarchical_2011}, though we note this is not always the case~\cite{vella_regimes_2018}.
In our system, the fold terminus occurs at a characteristic polar angle, and the amount of material stored in each fold grows further from the apex of the sphere in order to accommodate the curvature of the underlying substrate (\Fig{fig_morphology}d). 
This type of fold also appears in skirts and other clothing, where it is called a `dart'.

While we robustly observe pronounced folds on large PS spheres, we find no evidence for smaller-scale wrinkling in the sheets.  
This can be predicted from the energy scaling (\Fig{fig_energy_scaling}): the cost to delaminate from the PS surface exceeds both folding and stretching energies ($E_\gamma > E_f, E_s$).


\section{Conclusions}
In this article, we focused on the ability of preassembled nanoparticle monolayer sheets to conform to a substrate composed of a lattice of larger spheres. 
With its local Gaussian curvature, $K$, which can be tuned by varying the sphere diameter, such a substrate serves as a model for arbitrary surface topographies.
In the presence of strong pinning to the substrate, the area mismatch between flat ($K = 0$) and spherical ($K > 0$) geometries triggers a competition between different deformation modes of the sheet, including delamination, bending, stretching, fracture, and folding.

Treating the sheets as homogeneous continuum material leads to a scaling picture which is consistent with the general trends of elastic deformation in our system. 
For comparison with experiments, we extracted the local strain tensor components from images of the sheets, where the nanoparticles served as distance markers. 
While this analysis was consistent with our general scaling picture, the details of plastic deformation are only captured if the discrete nature of the sheets is taken into account, allowing changes in the number of nearest neighbors for individual particles. 
By tracking the onset of strain-induced dislocations within the sheets, we are able to explain deviations from the continuum predictions, which are found when the sheets are conformed to substrates with small $D$, corresponding to regions of large $K$.

The observed morphologies for the stamped sheets highlight the remarkable ability of nanoparticle monolayers to cope with strain through a combination of elastic and plastic deformations. 
This material contrasts with other thin sheets such as paper, mylar, or graphene, which lack a similar mechanism for generating particle dislocations.
We note that if the material properties of our sheets were tuned by changing the gold nanoparticle size, changing the ligand length, or functionalizing the ligands, a different sequence of morphological regimes could emerge as the substrate sphere size varies (Supplementary Information~\Fig{fig:altscaling}).

There is currently much interest in creating functional materials by stacking ultrathin, essentially 2D layers with different electronic or optical properties~\cite{androulidakis_tailoring_2018,kang_layer-by-layer_2017}.
So far, such stacking has been limited to flat substrates, where it is relatively easy to obtain good interfaces between successively deposited layers. 
In this regard, the ability of nanoparticle sheets to comply and conform opens up new possibilities for creating stacked layers with well-controlled interfaces also on more complex substrate topographies.



%

\section*{Acknowledgements}
We thank Anton Souslov, Vincenzo Vitelli, and William Irvine for useful discussions.
This work was supported by the Office of Naval Research under award ONR-N00014-17-1-2342 and by the National Science Foundation under award DMR-1508110. 
Additional support was provided by the University of Chicago Materials Research Science and Engineering Center, which is funded by National Science Foundation under award number DMR-1420709. 
Use of the Center for Nanoscale Materials was supported by the U.S. Department of Energy, Office of Science, Office of Basic Energy Sciences, under Contract No. DE-AC02-06CH11357.


%


\newcommand{\ds}{\displaystyle}
\newcommand{\sm}{\sum_{n=1}^\infty}
\newcommand{\be}{\begin{enumerate}}
\newcommand{\ee}{\end{enumerate}}
\newcommand{\bi}{\begin{itemize}}
\newcommand{\ei}{\end{itemize}}
\newcommand{\GG}{\mathcal{G}}
\newcommand{\CC}{\mathbb{C}}
\newcommand{\ZZ}{\mathbb{Z}}
\newcommand{\RR}{\mathbb{R}}
\newcommand{\NN}{\mathbb{N}}
\newcommand{\qq}{\mathbb{Q}}

\newcommand{\dif}{\mathrm{d}}

\newcommand{\AAA}{\mathbf{A}}
\newcommand{\FF}{\mathbf{F}}
\newcommand{\aaa}{\mathbf{a}}

\newcommand{\eee}{\mathbf{e}}
\newcommand{\mm}{\mathbf{m}}
\newcommand{\nn}{\mathbf{n}}
\newcommand{\pp}{\mathbf{p}}
\newcommand{\qqq}{\mathbf{q}}
\newcommand{\rr}{\mathbf{r}}
\newcommand{\sss}{\mathbf{s}}
\newcommand{\ttt}{\mathbf{t}}
\newcommand{\vv}{\mathbf{v}}
\newcommand{\ww}{\mathbf{w}}
\newcommand{\xx}{\mathbf{x}}
\newcommand{\zz}{\mathbf{z}}
\newcommand{\ttau}{\bm{ \tau}}
\newcommand{\om}{\omega}
\newcommand{\omm}{\boldsymbol \omega}
\newcommand{\Omm}{\boldsymbol \Omega}
\newcommand{\x}{\times}
\newcommand{\curl}{\nabla \times}
\newcommand{\dvr}{\nabla \cdot}
\newcommand{\grad}{\nabla}
\newcommand{\lap}{\nabla^2}
\newcommand{\pt}{\partial}
\newcommand{\gf}{\frac{1}{\Delta_{\xx \xx'}}}
\newcommand{\cross}{\times}
\newcommand{\trm}{\textrm}
\newcommand{\pphi}{\boldsymbol \varphi}
\newcommand{\EE}{\mathbf{E}}
\newcommand{\BB}{\mathbf{B}}
\newcommand{\JJ}{\mathbf{J}}
\newcommand{\KK}{\mathbf{K}}
\newcommand{\LL}{\mathbf{L}}
\newcommand{\II}{\mathbf{I}}
\newcommand{\MM}{\mathbf{M}}
\newcommand{\NNN}{\mathbf{N}}
\newcommand{\OO}{\mathbf{O}}
\newcommand{\PP}{\mathbf{P}}
\newcommand{\SSS}{\mathbf{S}}
\newcommand{\XX}{\mathbf{X}}
\newcommand{\YY}{\mathbf{Y}}
\newcommand{\ZZZ}{\mathbf{Z}}
\newcommand{\laplace}{\nabla^2}
\newcommand{\ldarrow}{\longleftrightarrow}
\newcommand{\darrow}{\leftrightarrow}
\newcommand{\bal}{\begin{align}}
\newcommand{\lab}{\end{align}}
\newcommand{\eps}{\varepsilon}
\newcommand{\eij}{\varepsilon_{ij}}
\newcommand{\exx}{\varepsilon_{xx}}
\newcommand{\exy}{\varepsilon_{xy}}
\newcommand{\eyy}{\varepsilon_{yy}}
\newcommand{\gij}{g_{ij}}
\newcommand{\dij}{\delta_{ij}}
\newcommand{\Evv}{E_{vv}}
\newcommand{\Ev}{E_{v}}
\newcommand{\Eu}{E_{u}}
\newcommand{\Fuv}{E_{uv}}
\newcommand{\Fv}{F_{v}}
\newcommand{\Fu}{F_{u}}
\newcommand{\Guu}{G_{uu}}
\newcommand{\Gv}{G_{v}}
\newcommand{\Gu}{G_{u}}
\newcommand{\sxx}{\sigma_{xx}}
\newcommand{\sxy}{\sigma_{xy}}
\newcommand{\syy}{\sigma_{yy}}
\newcommand{\ux}{u_x}
\newcommand{\uy}{u_y}
\newcommand{\px}{\partial_x}
\newcommand{\py}{\partial_y}
\newcommand{\ab}{\alpha \beta}
\newcommand{\db}{,\beta}
\newcommand{\da}{,\alpha}

\newcommand{\ti}{\textrm{I}}
\newcommand{\tii}{\textrm{II}}
\newcommand{\tiii}{\textrm{III}}
\newcommand{\tic}{\textrm{Ic}}
\newcommand{\ki}{K_{\textrm{I}}}
\newcommand{\kii}{K_{\textrm{II}}}
\newcommand{\bth}{\bar{\theta}}

\newcommand{\one}{\mathbf{\hat{1}}}
\newcommand{\done}{\mathbf{\dot{\hat{1}}}}
\newcommand{\two}{\mathbf{\hat{2}}}
\newcommand{\dtwo}{\mathbf{\dot{\hat{2}}}}
\newcommand{\three}{\mathbf{\hat{3}}}
\newcommand{\dthree}{\mathbf{\dot{\hat{3}}}}

\newcommand{\phidot}{\dot{\phi}}
\newcommand{\thetadot}{\dot{\theta}}
\newcommand{\psidot}{\dot{\psi}}
\newcommand{\Xdot}{\dot{X}}
\newcommand{\Ydot}{\dot{Y}}

\newcommand{\xhat}{\mathbf{\hat{x}}}
\newcommand{\yhat}{\mathbf{\hat{y}}}
\newcommand{\zhat}{\mathbf{\hat{z}}}
\newcommand{\Xhat}{\mathbf{\hat{X}}}
\newcommand{\Yhat}{\mathbf{\hat{Y}}}
\newcommand{\Zhat}{\mathbf{\hat{Z}}}
\newcommand{\deltahat}{\boldsymbol {\hat{\delta}}}
\newcommand{\phihat}{\boldsymbol {\hat{\phi}}}
\newcommand{\nhat}{\mathbf{\hat{n}}}
\newcommand{\ddelta}{\boldsymbol\delta}

\newcommand{\bilaplace}{\nabla^4}
\newcommand{\Tr}{\textrm{Tr}}
\newcommand{\kkappa}{\boldsymbol \kappa}
\newcommand{\ssigma}{\boldsymbol \sigma}

\cleardoublepage
\begin{titlepage}
   \vspace*{\stretch{1.0}}
   \begin{center}
      \large\textbf{Supplementary Information for \\ `Conforming nanoparticle sheets to surfaces with Gaussian curvature'}\\
   \end{center}
   \vspace*{\stretch{3.0}}
\end{titlepage}
\renewcommand{\thefigure}{S\arabic{figure}}
\renewcommand\theequation{S\arabic{equation}}
\setcounter{figure}{0} 
\setcounter{equation}{0}

\section{Bending}
In~\Fig{fig_energy_scaling} of the main text, we omit linear and sublinear dependences on the polar angle, $\theta$, for clarity. 
As a result, the bending energies for different polar angles (blue, gray, and orange dashed lines) are shown to lie atop each other.
Here we note that we expect some dependence of the bending energy density $\mathcal{E}_\textrm{b}$ on polar angle, though this should appear as a subleading, quadratic correction to the bending energy on the apex of a PS sphere.
The leading behavior is therefore $E_\textrm{b} \sim (B / D)\, \delta r \sin \theta $, where $B$ is the bending modulus of the sheet and $D$ is the diameter of the sphere.

The two-dimensional bending energy density of a thin plate in plane stress is~\cite{landau_chapter_1986}
\begin{multline}\label{eq_bending}
\mathcal{E}_{\textrm{b}} = \frac{B}{2} \left[ \left(\nabla^2 \zeta 
\right)^2 \right.\\
\left.
+ 2 (1 - \nu) 
\left\{
\left(\frac{\partial^2 \zeta}{\partial x \partial y}
\right)^2 -
\frac{\partial^2 \zeta}{\partial x^2}
\frac{\partial^2 \zeta}{\partial y^2}
\right\}
\right],
\end{multline}
where $\zeta(x,y)$ is the out-of-plane displacement of the plate and $B$ is the bending modulus.
Taylor expanding around $\theta = 0$, the energy density evaluates to
\begin{equation}
\begin{split}
\mathcal{E}_\textrm{b} = \frac{4B}{D^2} \bigg[&
 (\nu +1) + 
2 (\nu +1) \theta^2   \\
 & + \frac{14 \nu +17}{6} \theta^4 +
\mathcal{O}\left(\theta^5\right)
 \bigg].
\end{split}
\end{equation}
Thus, we expect the bending energy of a membrane to increase with polar angle.
This analysis neglects the presence of neighboring spheres, which would further affect the $\theta$ dependence, particularly at large $\theta$, where the small deflection assumption and the validity of this expression for bending energy density break down.


\section{Stretching}


\subsection{Definitions of stretching energy, strain, and stress}

Assuming locally in-plane displacements $\uu(r,\phi) = u_r(r,\phi)\hat{\bf{r}} + u_\phi(r,\phi) \hat{\boldsymbol{\phi}}$, we have strains \cite{landau_chapter_1986}
\begin{align}
&\varepsilon_{rr} = \partial_r u_r \\
&\varepsilon_{\phi\phi} = \frac{1}{r} \partial_\phi u_\phi + \frac{1}{r} u_r\\
&\varepsilon_{r\phi} = \frac{1}{2} \left( \frac{1}{r} \partial_\phi u_r + \partial_r u_\phi  \right).
\end{align}
When the out-of-plane displacements are included, the expressions for strain become
\begin{align}
&\varepsilon_{rr} = \partial_r u_r + \frac{1}{2}(\partial_r \zeta)^2 \\
&\varepsilon_{\phi\phi} = \frac{1}{r} \partial_\phi u_\phi + \frac{1}{r} u_r + \frac{1}{2r^2}(\partial_\phi \zeta)^2\\
&\varepsilon_{r\phi} = \frac{1}{2} \left( \frac{1}{r} \partial_\phi u_r + \partial_r u_\phi  + \frac{1}{r} \partial_r \zeta \partial_\phi \zeta \right).
\end{align}
These strains are related to the stress via
\begin{align}
&\sigma_{rr} = \frac{Y}{1-\nu^2} (\varepsilon_{rr} + \nu \varepsilon_{\phi \phi}) \\
&\sigma_{\phi \phi} = \frac{Y}{1-\nu^2} (\varepsilon_{\phi\phi} + \nu \varepsilon_{rr}) \\
&\sigma_{r\phi} = \frac{Y}{1+\nu} \varepsilon_{r\phi},
\end{align}
where $Y = Et$ is the stiffness.

The stretching energy density, $\mathcal{E}_{s} = \frac{1}{2} \sigma_{ij} \varepsilon_{ij}$, takes the plane stress form
\begin{multline}
\mathcal{E}_{s} = \frac{Y}{1-\nu^2}\left(\frac{\varepsilon_{rr}^2 + \varepsilon_{\phi\phi}^2}{2} 
+ \nu \varepsilon_{rr} \varepsilon_{\phi\phi} 
\right) 
\\
+ \frac{2Y}{1+\nu}\varepsilon_{r\phi}^2.
\end{multline}
Since $\varepsilon_{r\phi}=0$ by symmetry on the sphere,
\begin{equation}
\mathcal{E}_{s} = \frac{Y}{1-\nu^2}\left(\frac{\varepsilon_{rr}^2 + \varepsilon_{\phi\phi}^2}{2} 
+ \nu \varepsilon_{rr} \varepsilon_{\phi\phi} 
\right).
\end{equation}


\begin{figure}
\includegraphics[width=\columnwidth]{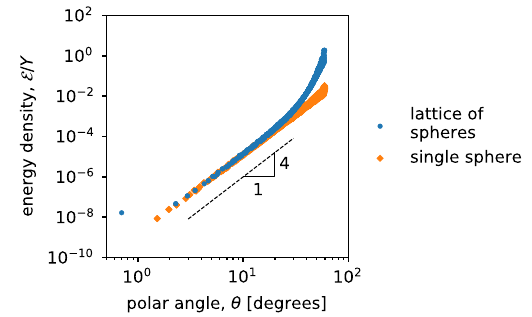}
\caption{
\textbf{Stretching energy in a spring network draped on a lattice of spheres with strong pinning.}
The energy density in a pinned sheet draped to a lattice of spheres grows as $\theta^4$. 
Only at moderately large polar angles ($\theta \gtrsim 25^\circ$) does the stretching energy in a sheet conforming to a triangular lattice of spheres (blue circles) diverge from the case of a single sphere (orange diamonds).
The quartic scaling with polar angle is exact in the absence of neighboring substrate spheres (orange diamonds).
Both spring networks were $100 \, a \times 100 \, a$ in extent, and the substrate sphere diameters were $40\,a $ and $60 \, a$ for the lattice and single sphere cases, respectively.
}
\label{fig_stretching_energy}
\end{figure}

\begin{figure}
\includegraphics[width=\columnwidth]{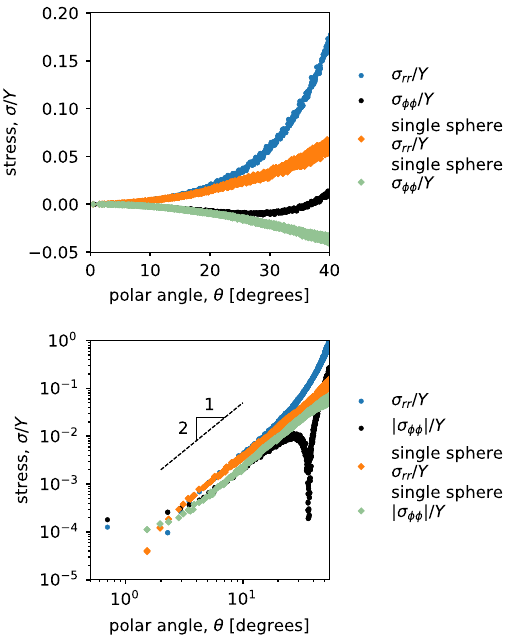}
\caption{
\textbf{In-plane stresses in a spring network draped on a lattice of spheres with strong pinning.}
The stress density in a pinned sheet draped to a lattice of spheres grows as $\theta^2$. 
Only at moderately large polar angles ($\theta \gtrsim 25^\circ$) does the stretching energy in a sheet conforming to a triangular lattice of spheres (blue circles) diverge from the case of a single sphere (orange diamonds). 
The quadratic scaling with polar angle is exact in the absence of neighboring substrate spheres (gray and orange diamonds for $\sigma_{rr}$ and $\sigma_{\phi\phi}$, respectively.)
The lattice dimensions are the same as in~\Fig{fig_stretching_energy}.
}
\label{fig_stress}
\end{figure}


\subsection{Sequential pinning gives $\mathcal{E}_s \sim Y \theta^4$}

\Fig{fig_stretching_energy} shows the stretching energy of an elastic spring network as a function of polar angle on the sphere.
We find the stretching energy density grows as $\mathcal{E}_s \sim Y \theta^4$ for modest polar angle.
Additionally, each component of the stress exhibits $\sigma \sim Y \theta^2$ scaling, particularly when only a single sphere is present as the substrate, as shown in~\Fig{fig_stress}. 
The presence of neighboring spheres in the substrate causes deviation from the power-law scaling in both energy density and stress for sufficiently large polar angles ($\theta \gtrsim 25^\circ$).
The quadratic scaling of the strain, $\varepsilon$, can likewise be seen in~\Fig{fig_strain_plots} of the main text.

The geometric frustration of the sheet on the spherical cap is the source of elastic energy in an annulus of the sheet that has not yet conformed to the sphere. 
In particular, let us consider the portion of the sheet near $\theta_a$ which is just about to adhere to the sphere, and is therefore about to become pinned in its current state of strain.
The strain at $\theta_a$ scales linearly with the integrated Gaussian curvature of the spherical cap:
$\varepsilon \sim \int_0^{R \theta_a} G \, r \, \textrm{d} r \sim \int_0^{R \theta_a} (1/R^2) \, r \, \textrm{d} r
\sim R^0 \theta_a^2$, where $R=D/2$ is the radius of the sphere~\cite{mitchell_fracture_2017,vitelli_crystallography_2006}.
This portion of the sheet is then frozen into a strain configuration that depends quadratically on the polar angle at which it conforms.
As a result, after many annuli have adhered, each corresponding to a ever-larger $\theta_a$, we expect $\varepsilon \sim \theta^2$.
Linear elasticity dictates that the stress scales similarly as well --- $\sigma \sim Y \varepsilon \sim Y \theta^2$, where $Y$ is the stiffness --- and thus the stretching energy density $\mathcal{E}_s =\frac{1}{2}\sigma\varepsilon \sim Y \theta^4$. 
This means that the stretching energy stored in an annulus is $E_s \sim Y  D \,  \delta r \,\theta^4 \sin \theta$, which for small $\theta$ gives $E_s \sim Y  D \, \delta r\, \theta^5$.
Sequential pinning of the nanoparticle sheet ensures that this is true irrespective of the maximum angle subtended by the sheet: the state of strain is frozen into the adhered portion, unable to respond elastically to additional pileup of strain at $\theta > \theta_a$.


\subsection{Case without pinning has different $\mathcal{E}_s$ scaling}
This analysis contrasts with the expectation for an equilibrated elastic sheet without pinning.
Without pinning, the energy density rearranges in such a way as to be non-monotonic in the polar angle $\theta$ on the sphere, with some sensitivity to the boundary conditions. 
The stress is greatest on the apex of a sphere without pinning, in stark contrast to the case with sequential pinning, for which the stress vanishes at the cap.
This difference highlights the distinct character of sequential adhesion to a substrate seen in our system.

Without pinning, we can solve for the strain energy by finding the stress and strain via 
\begin{equation} \label{bilaplacian}
\frac{1}{Y} \nabla^4 \chi(r) = - G = -\frac{1}{R^2},
\end{equation}
where $\chi$ is the Airy stress function given by $\sigma_{ij} = \varepsilon_{il} \varepsilon_{jk} \partial_l \partial_k \chi $ and where, as before, $R=D/2$ is the radius of the sphere.
Solving~\Eq{bilaplacian} for the energy density in a circular sheet of radius $W$ equilibrated to a spherical cap results in 
\begin{align}
\nonumber 
\mathcal{E}_s (r) = & 
\frac{G^2 Y}{256} \left[(5-3 \nu ) r^4 - 4 (1 - \nu) r^2 W^2  \right.
\\ & \left. + (1 - \nu ) W^4\right] 
\\ & \nonumber
+ 
\frac{ G P (\nu - 1)}{8 Y} \left(2 r^2-W^2\right)
+ \frac{T^2 ( 1-\nu)}{Y},
\end{align}
where $r$ is the radial coordinate of the polar coordinate system on the apex of the sphere, and the apex is assumed to coincide with the center of the circular sheet.
Here, $T=\sigma_{rr}(r=W)$ is the radial stress at the boundary.
If we set $T=0$ for the moment to look only at the effects of curvature, for small polar angles $\theta \approx r\sqrt{G} $, the energy density \textit{decreases} with polar angle in a quadratic correction:
\begin{equation}
\mathcal{E}_s \approx \frac{G^2 Y (1-\nu)}{256} \left[ W^4 
- \frac{4  W^2}{G} \theta^2
+ \mathcal{O}(\theta^4) \right].
\end{equation} 
If we set $W=R \theta_{\textrm{max}} = D \theta_{\textrm{max}} /2$, with $\theta_{\textrm{max}}$ fixed, the leading term shows that $\mathcal{E}_s \sim  Y D^0\theta^0$.
This behavior contrasts with the case with pinning studied elsewhere in this article.
We note, however, that the \textit{total} stretching energy in the entirety of a spherical cap conformed to a sphere, with or without pinning, has the \textit{same} scaling: $E_s^{\textrm{tot}} \sim Y D^2 \theta_{\textrm{max}}^6 $, where $\theta_{\textrm{max}}$ is the maximum angle at the edge of the sheet.
\begin{figure}
\includegraphics[width=\columnwidth]{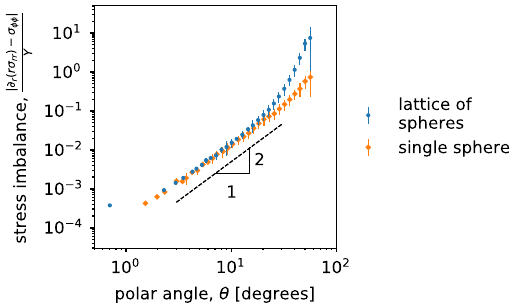}
\caption{
\textbf{Adhesion enables an in-plane stress imbalance to the elastic membrane.}
Through adhesion to the substrate, there is a residual force imbalance in the stretching of a simulated triangular spring network.
The quadratic scaling with polar angle is exact in the absence of neighboring substrate spheres (orange diamonds).
Both spring networks were $100 \, a \times 100 \, a$ in extent, and the substrate sphere diameters were $40\,a $ and $60 \, a$ for the lattice and single sphere cases, respectively.
}
\label{fig_imbalance}
\end{figure}


\begin{figure}
\includegraphics[width=\columnwidth]{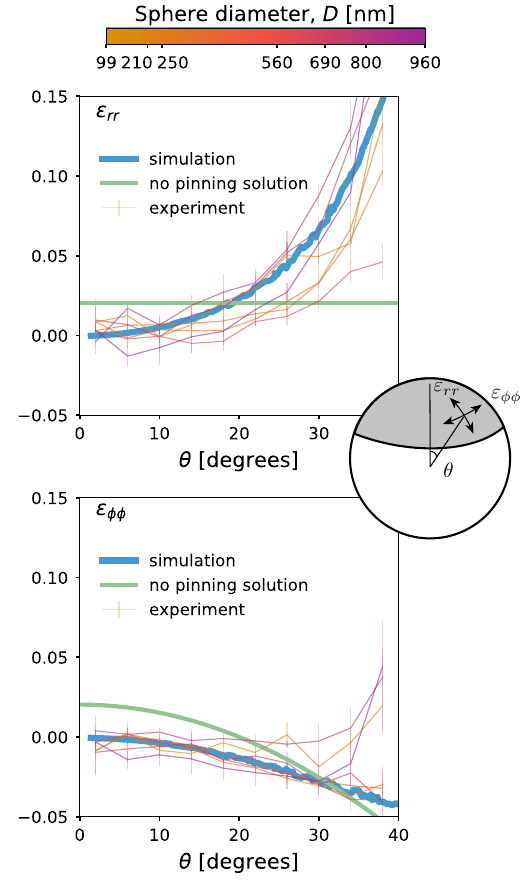}
\caption{
\textbf{Strong pinning to the substrate is necessary for qualitative agreement with experiments.}
The analytic solution in the case with no adhesion, given by the green curve, differs qualitatively from the simulation (blue curve) and experimental results (transparent orange and purple data). 
}
\label{fig_nopinning}
\end{figure}

\begin{figure}
\includegraphics[width=\columnwidth]{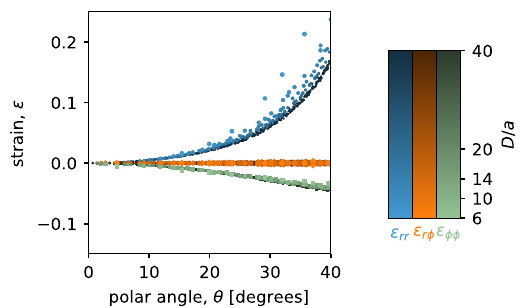}
\caption{
\textbf{Finite size effects in the energetics of draped spring networks.}
Spring networks with lattice spacing $a$ were draped over seven spheres in a triangular closed packed arrangement, as in~\Fig{fig_setup_sim} of the main text. 
The resulting strains depend only weakly on the ratio of sphere size to lattice spacing, $D/a$, so that the data coincide for all but the smallest values of $D/a$.
While the shear and azimuthal strains are nearly unaffected by the size of the lattice, the radial strain begins to diverge significantly around $D/a \sim 10$. 
}
\label{fig_finsize}
\end{figure}

\subsection{Influence of adhesion}
\Fig{fig_imbalance} shows that for modest polar angles, the in-plane stress imbalance
\begin{equation} 
\mathcal{I} \equiv \partial_r \left(r \sigma_{rr}\right) - \sigma_{\phi \phi}
\end{equation}
grows quadratically in simulations of spring networks draping to spheres.
Without adhesion, this quantity would vanish in equilibrium.
We checked that the residual force imbalance is scale-independent for sufficiently large substrate sphere sizes ($D/a \gtrsim 10$).

\subsection{Case without pinning does not agree with experiment}

If adhesion is not included, then the resulting strain field contrasts with the results from simulations, as shown in~\Fig{fig_nopinning}. 
The strain fields in this case are
\begin{align}
\varepsilon_{rr} &= \frac{1}{16} \left[ (3 \nu -1) \theta^2 +4 (1 - \nu) \frac{W^2}{D^2} \right]  
\\
\varepsilon_{\phi\phi} &= \frac{1}{16} \left[
(\nu - 3)\theta^2 + 4(1 - \nu) \frac{W^2 }{D^2}
\right],
\end{align}
where $W$ is the width of the sheet from the cap to the periphery.
We assume the radial stress vanishes at the boundary for simplicity ($T=0$), but we note that changing $T$ simply adds a constant to each strain component.
In~\Fig{fig_nopinning}, $W$ was taken to be the radius of the sphere, $D/2$, times 40$^\circ$ --- approximately where cracks appear in~\Fig{fig_morphology}c of the main text.
The qualitative differences in elastic response shown in~\Fig{fig_nopinning} highlight the importance of adhesion in determining the mechanical response and monolayer morphology.

\subsection{Finite size effects in draped spring networks}
We investigated the effects of finite size in the simulations of spring networks with respect to the substrate sphere size.
Spring networks with lattice spacing $a$ are draped over seven spheres in a triangular closed packed arrangement, as in~\Fig{fig_setup_sim} of the main text. 
The resulting strains depend weakly on the ratio of sphere size to lattice spacing, $D/a$.
While the shear and azimuthal strains are nearly unaffected by the size of the lattice down to values of $D/a \sim 6$, the radial strain begins to diverge significantly around $D/a \sim 10$.
This is reminiscent of previous work on nanoparticle membranes~\cite{Wang_strong_2015}, where the influence of the discrete lattice becomes significant for systems with a characteristic size of $\sim 10 \, a$.


\begin{figure}
    \centering
    \includegraphics[width=\columnwidth]{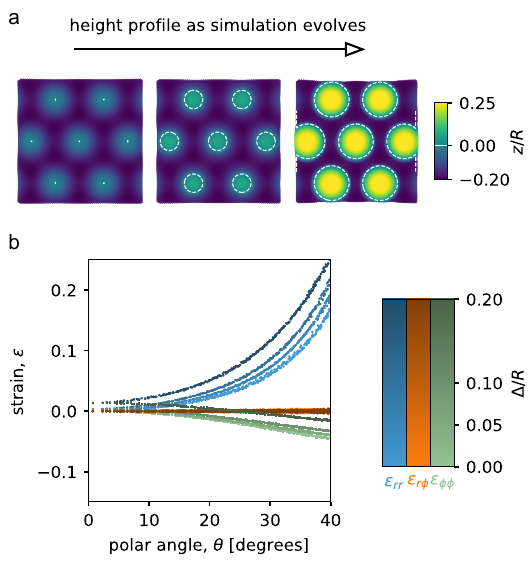}
    \caption{
    \textbf{Introducing a model indentation that penetrates the interstices of the lattice of substrate spheres changes the resulting nanoparticle sheet strain fields only slightly.}
    \textit{(a)} The spring network relaxes on a corrugated surface, which is lowered onto a lattice of seven spheres. 
    The white dashed curves mark the $z=0$ point, which is identified with the maximum height of the corrugated surface, $V(\xx)$, in the absence of substrate spheres. 
    The spheres then protrude from this surface as the simulation evolves, and the spring network relaxes on a surface $z=\textrm{max}\left\{V(\xx), z_{\textrm{spheres}}\right\}$. 
    Here the corrugation has a maximum depth of $\Delta = 0.2 R$, where $R$ is the radius of each substrate sphere.
    \textit{(b)} Changing the indentation depth results in only modest changes in the strain configuration of the pinned sheet at the end of the simulation. 
    Each value of maximum indentation depth, $\Delta$, corresponds to each shade of blue ($\varepsilon_{rr}$), orange ($\varepsilon_{r\phi}$), and green ($\varepsilon_{\phi\phi}$) curves.
    As $\Delta$ (here normalized by the substrate sphere radius $R$) increases from simulation to simulation, the qualitative behavior of the strains remains relatively unchanged.
    }
    \label{fig:interstices}
\end{figure}

\subsection{Sensitivity to conformation geometry}

In simulations reported so far, we have used a sheet geometry in which each nanoparticle lies either in an $xy$ plane at a decreasing $z$ position or on a sphere, whichever has a greater value of $z$ coordinate. 
However, we do not expect that the nanoparticle sheet will be truly flat in the interstices of the PS spheres in our experiments. 
For the simulations presented earlier, the sheet is equilibrated in each timestep on a surface defined by $z=\textrm{max}\{z_{\textrm{plane}}, z_{\textrm{spheres}}\}$ --- that is, each node of the network may reside on either a substrate sphere or in a plane which is lowered incrementally at each time step. 
At the end of each time step, nodes that reside on a substrate sphere are pinned to that location permanently.

\Fig{fig:interstices} shows that introducing a model indentation between substrate spheres elevates the observed strains of the final, pinned nanoparticle sheet.
In~\Fig{fig:interstices}b, each set of curves for $\varepsilon_{rr}$, $\varepsilon_{r\phi}$, and $\varepsilon_{\phi\phi}$ corresponds to a new simulation in which the spring network is iteratively stamped onto a lattice of spheres while conformed not to a plane, but to a corrugated surface with indentations penetrating the interstices of the substrate spheres (\Fig{fig:interstices}b). 
The networks are relaxed on a surface defined by $z=\textrm{max}\left\{V(\xx), z_{\textrm{spheres}}\right\}$, where
\begin{multline}
    V(\xx) = \Delta \big( \cos(\mathbf{b}_1 \cdot \xx)  
     + \cos(\mathbf{b}_2 \cdot \xx) 
     \\
     + \cos\left[ \left(\mathbf{b}_1  + \mathbf{b}_2 \right) \cdot \xx \right] \big),
\end{multline}
where $\mathbf{b}_1$ and $\mathbf{b}_2$ are the reciprocal lattice vectors of the honeycomb lattice defined by the position of the interstices.
This corrugated surface changes the angle of contact between the spring lattice and the substrate spheres and acts as a source of strain in the interstices.
While the qualitative strain behavior of the resulting pinned spring lattices remains largely unaltered, the radial and azimuthal strains grow with indentation depth, $\Delta$.
Future work could implement a more realistic boundary condition for downward pressure on the sheet, as $V(\xx)$ is a highly simplified surface.

\begin{figure}
    \centering
    \includegraphics[width=\columnwidth]{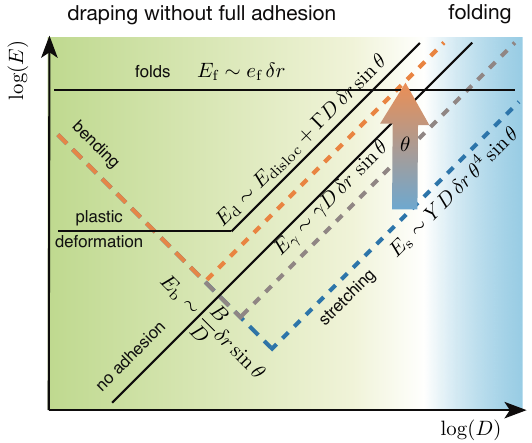}
    \caption{
    \textbf{A nanoparticle sheet's material properties could be altered such that $E_\textrm{d} > E_\gamma$ for all sphere sizes $D$.}
For this ordering of competing energy scales, there is only a transition from incomplete adhesion at small sphere sizes to folding at large sphere sizes.}
\label{fig:altscaling}
\end{figure}

\section{Other possible scaling diagrams}
In section 3 (entitled `Energy Scaling') and~\Fig{fig_energy_scaling} of the main text, we presented a competition of energy scales that captures the observed behavior of our nanoparticle sheets.
We note that, in a different material, the energetic cost of plastic deformation, captured via the phenomenological factor $\Gamma$, could be much larger than the adhesion energy, $\gamma$. In this case, the plastic deformation regime might be absent if no crossover between $E_\gamma$ and $E_\textrm{d}$ occurs.
This situation is illustrated in~\Fig{fig:altscaling}.

\end{document}